\documentclass[11pt,a4]{article}

\usepackage{amssymb}
\usepackage{amsmath}
\usepackage{graphicx}
\usepackage[numbers,compress]{natbib}
\begin{document}

\title{Rest frame vacua of massive Klein-Gordon fields on spatially flat FLRW spacetimes}

\author{Ion I. Cot\u aescu \footnote{e-mail: i.cotaescu@e-uvt.ro}\\
{\small \it West University of Timi\c soara,}\\{\small \it V. P\^ arvan Ave. 4, RO-300223, Timi\c soara, Romania}}

\maketitle

\begin{abstract}
We propose a method of projecting the quantum states from a state space of a given geometry into another state space generated by a different geometry, taking care on the correct normalization which is crucial in interpreting the quantum theory.  Thanks to this method we can define on any spatially flat FLRW spacetime  states in which genuine Minkowskian parameters are measured. We use these Minkowskian states for separating the frequencies in the rest frames of the massive scalar particles defining thus the scalar rest frame vacuum. We show that this vacuum is stable on the de Sitter expanding universe where the energy is conserved. In contrast, on a spatially flat FLRW spacetime with a Milne-type scale factor this vacuum results to be dynamic,  corresponding to a time-dependent rest energy interpreted as an effective mass. This dynamic vacuum give rise to a cosmological particle creation which is significant only in the early Milne-type universe considered here. Some interesting features of this new effect are pointed out in a brief  analysis.

\end{abstract}

\section{Introduction}

One of the most studied semi-classical approaches is the cosmological particle creation (c.p.c.) in which one considers only the interaction of free particles, seen as quantum perturbations, with the gravity of a time-dependent curved background whose evolution remains unaffected by this interaction.  This approach consists in separating the particle and antiparticle quantum modes at different epochs of  the evolving spacetime  determining thus different bases of the state space each one corresponding to a specific vacuum. These may be related among themselves through  Bogolyubov transformations whose transition coefficients may point out the cosmological particle creation generating particle or antiparticle thermal baths \cite{P1,P2,GM,Zel,ZelS,DW,GMM,U1,Br,GMM1,U2}.  A special attention was paid to the scalar field on the de Sitter spacetime \cite{Nach,CT,BODU,T,Csc,Pascu,Csc1} involved in many studies of c.p.c. \cite{h1,BuD,h2,h3,h4,h5,h6,h7,h8,h9,h10,h11,Ambrus,h12,h13}.

The main task here is just the criterion of separating the frequencies defining the particle and antiparticle modes and, implicitly, the current vacuum at a given time.  The principal method used so far is to focus mainly on the asymptotic states whose behavior is similar to the usual Minkowskian particle and antiparticle mode functions. In this manner one may choose $in$ and $out$ states whose frequencies are separated as in the flat case defining thus the adiabatic vacua as, for example, the Bunch-Davies one \cite{BuD} largely used in applications. 

On the other hand, recently we proposed the rest frame vacuum (r.f.v.) of the massive Dirac field on $(1+3)$-dimensional spatially flat Friedmann-Lema\^ itre-Robertson-Walker (FLRW) spacetimes \cite{Crfv}. We started with the observation that in the rest frame, where the particle momentum vanishes,  the solutions of the Dirac equation on any FLRW spacetime  have a Minkowskian behavior regardless the time evolution of the background. Thus we can separate the frequencies as in special relativity obtaining a  time-independent vacuum but which is different from the Bunch-Davies one we used so far \cite{CdS,CQED}. Note that the r.f.v. can be defined only for massive particles since the massless ones do not have rest frames.     

The next step might be the generalization of the r.f.v. to the Klein-Gordon and Proca fields seen as perturbations on the mentioned manifolds. Unfortunately, here we face with a serious difficulty since, in contrast with the Dirac field, the rest mode functions of these bosonic fields do not have the Minkowskian forms we need for defining the r.f.v. in a natural manner. Nevertheless, since  our concept of particle and antiparticle comes from the Minkowskian quantum field theory, we  are forced to impose the Minkowskian forms to the rest mode functions even though this is possible only at a given time, assuming that the time-dependent rest energy represents  a {\em dynamical} effective mass.  

Another challenge is to solve the ambiguity related to the normalization of these Minkowskian states, which can be done in two manners, either with respect to the scalar product of the curved manifold or by using the Minkowskian scalar product. This difficulty can be avoided since the state spaces are separable Hilbert spaces which are isometric among themselves such that  we may look for an appropriate method of mapping  the state spaces produced in different geometries into a unique one where we may calculate transition coefficient between these states.  In what follows we would like to concentrate on these problems proposing a method of defining well-normalized Minkowskian  rest mode functions on any spatially flat FLRW spacetime. These will help us to define at any moment  the bosonic  time-dependent r.f.v. associated to a time-dependent dynamical mass. When this is time-independent we say that the r.f.v. is {\em stable}.   For concrete calculations we restrict ourselves to the massive Klein-Gordon field minimally coupled to the background gravity.  

In our approach the r.f.v. of the scalar field are stable only on the FLRW manifolds where the energy is conserved as in the de Sitter expanding universe. In other FRLW manifolds we  meet c.p.c. processes that may be studied deriving the Bogolyubov coefficients between  states whose vacua are defined at two arbitrary moments,  obtaining thus information about the time behavior of the c.p.c. in any FLRW geometry. For illustrating how our method works we give two examples, the stable r.f.v. on the de Sitter expanding universe and, for the first time, we present an example of time-dependent r.f.v. on a spatially flat FLRW with a Milne-type scale factor.  On this last manifold we study the c.p.c. at finite times obtaining probabilities and rates which depend exclusively on the moments when the particle is prepared and then measured. Note that our results are different from other attempts of studying c.p.c. at finite times \cite{Ambrus} where the vacuum depends, in addition, on momentum. 

We start in the next section presenting our basic assumptions concerning the scalar quantum modes prepared or measured by a global apparatus on a curved manifold and shoving how the state space can be mapped into a Minkowskian one. The next section is devoted to the spatially flat FLRW spacetimes where we propose a concrete method of defining Minkowskian rest states, correctly normalized at a given arbitrary time, regardless the time evolution of the background geometry.  By using such states we define the r.f.v. showing that these vacua are stable only on the FLRW manifolds where the energy is conserved. The  sections IV and V are devoted to the mentioned examples, the stable r.f.v. on de Sitter expanding universe and respectively the time-dependent one on the Milne-type universe. In this last section we study the c.p.c. produced by the vacuum instability deriving the Bogolyubov coefficients between two bases of mode functions whose frequency separation was performed at two different moments. Some physical consequences are briefly discussed based on a brief analytical and graphical analysis.   In the last section we present our concluding remarks.

\section{Minkowskian scalar modes}

Respecting {\em ad litteram} the principles of the quantum theory,  we assume that the quantum states on any  curved manifold, $(M,g)$, are prepared or measured by a global apparatus represented by the algebra of the quantum observables, i. e. the Hermitian operators defined globally as vector fields on the whole manifold or on a portion with an independent physical meaning, as in the case of the de Sitter expanding universe.  The operators proportional with the Killing vector fields are conserved, commuting with the operator of the field equation. Our global apparatus prepares quantum modes whose mode functions are common eigenfunctions of a system of commuting conserved operators (s.c.c.o.) $\{{\cal E}_{KG},A,B,...\}$ which includes the operator of the field equation ${\cal E}_{KG}$. In addition, these mode functions are supposed to be normalized with respect to a specific relativistic scalar product on $(M,g)$. 

In general, the s.c.c.o. determining the quantum modes is not complete such that the mode functions remain with  some integration constants which depend on the separation of  the positive and negative frequencies defining the vacuum. Another possible manner of setting these constants is by defining the  modes on $(M,g)$ in which one measures the parameters corresponding to another geometry $(\hat M,\hat g)$, according to the method we present in what follows. 

We start with the $(1+3)$-dimensional curved manifold $(M,g)$, supposed to be local Minkowskian, where we consider a local chart $\{x\}$  of coordinates $x^{\mu}$ (labeled by natural indices $\alpha, ...\mu,...=0,1,2,3$) with $x^0=t$ and arbitrary space coordinates. The scalar field, $\Phi: M\to {\Bbb C}$, of mass $m$, minimally coupled to the gravity of  $(M,g)$,  satisfies the Klein Gordon equation ${\cal E}_{KG}\Phi=m^2 \Phi$ whose operator is defined as
\begin{equation}\label{KG}
{\cal E}_{KG}=-\frac{1}{\sqrt{g}}\,\partial_{\mu}  \sqrt{g}\,
g^{\mu\nu}\partial_{\nu}\,,\quad g=|{\rm det} g_{\mu\nu}|\,.
\end{equation}
The solution of this equation may be expanded in terms of the mode functions $f_{\alpha}\equiv f_{a,b,...}$ which satisfy the Klein-Gordon equation and the eigenvalues ones $Af_{a,b...}=a f_{a,b,...},\,Bf_{a,b...}=b f_{a,b,...},...$,  determining these functions partially or completely when the s.c.c.o. is complete. When the eigenvalues $\alpha\equiv\{a,b,...\}$ are of continuous spectra this expansion reds
\begin{equation}
\Phi(x)=\int d\alpha\, \left[f_{\alpha}(x){\frak a}(\alpha)+f^*_{\alpha}(x){\frak a}^c(\alpha)^{\dagger}\right]\,,
\end{equation}
where the particle, ${\frak a}, {\frak a}^{\dagger}$, and antiparticle, ${\frak a}^c, {\frak a}^{c\,\dagger}$, field operators must satisfy the canonical bosonic commutation relations \cite{BD}.

The mode functions, $f\in{\cal K}$, behave as tempered distributions or square integrable functions with respect to the indefinite Hermitian form  
\begin{eqnarray}\label{SPgen}
\langle f,f'\rangle_M&=&i\int_{\Sigma} d\sigma^{\mu}\sqrt{g}\,  f^*\stackrel{\leftrightarrow}{\partial}_{\mu}f' \nonumber\\
&=&i\int_{{\Bbb R}^3} d^3x\, g^{00}\sqrt{g}\,  f^*\stackrel{\leftrightarrow}{\partial}_{t}f' \, \in {\Bbb C} \,,
\end{eqnarray}
written with  the notation $f\stackrel{\leftrightarrow}{\partial}f'= f \partial f' -f'\partial f$. This is the relativistic scalar product giving the 'squared norms' $\langle f,f\rangle_M$ of the square integrable functions $f\in {\cal H}\subset{\cal K}$ which may have any sign  splitting  the space ${\cal K}$ as
\begin{equation}
f \in \left\{ 
\begin{array}{lll}
{\cal H}_+\subset{\cal K}_+& {\rm if}& \langle f,f\rangle_M>0\,,\\
{\cal H}_0\subset{\cal K}_0& {\rm if}& \langle f,f\rangle_M=0\,,\\
{\cal H}_-\subset{\cal K}_-& {\rm if}& \langle f,f\rangle_M<0\,.\\
\end{array}
\right.
\end{equation} 
From the physical point of view  the mode functions of  ${\cal K}_{\pm}$ are of positive/negative frequencies while those of  ${\cal K}_0$ do not  have a physical meaning. For any $f\in {\cal K}_+$ we have $f^*\in {\cal K}_-$ so that $\langle f^*, f^*\rangle_M =-\langle f, f\rangle_M$ but whether  $f^*=f$ then  $f\in {\cal  F}_0$, since $\langle f, f\rangle_M=0$.   In fact,  ${\cal H}$ is a Krein space while ${\cal K}_{\pm}$ are the spaces of  tempered distributions of the Hilbertian triads associated to the Hilbert spaces ${\cal H}_{\pm}$ equipped with the scalar products $\pm \langle~,~\rangle_M$. 

A complete system of orthonormal mode functions, $\{f_{\alpha}\}_{\alpha\in I}\subset{\cal K}_+$ forms a (generalized) basis of positive frequencies in ${\cal K}_+$ related to the negative frequencies one, $\{f^*_{\alpha}\}_{\alpha\in I}\subset{\cal K}_-$. In this manner one defines a frequencies separation  associated to  a specific vacuum state of the Fock space. It is known that  two different bases define different vacuum states when these are related among themselves through a non-trivial Bogolyubov transformation that mixes the positive and negative frequency modes. Otherwise the vacuum state remains stable.

Furthermore, we consider  another manifold $(\hat M,\hat g)$ whose local chart $\{\hat x\}$ is defined on the {\em same} domain of the flat model as the chart $\{x\}$ of $(M,g)$. This  means that there exists the coordinate transformation $\hat x=\chi(x)$ allowing us to relate the set ${\cal K}$ discussed above to the set $\hat{\cal K}$ of the scalar mode functions on $(\hat M, \hat g)$ equipped with the Hermitian form $\langle\,,\,\rangle_{\hat M}$, defined as in Eq. (\ref{SPgen}). We observe that the physical parts of the sets $\hat{\cal K}$ and ${\cal K}$ are separable Hilbert spaces between which we can define the isometry $\mu: {\cal H}_+\to \hat{\cal H}_+$  which satisfies
\begin{equation}
\langle \mu(f), \mu(f')\rangle_{\hat M}=\langle f, f'\rangle_M\,.
\end{equation}
Then for any normalized mode functions $f_{\alpha}\in {\cal H}_+$ and $\hat f_{\beta}\in \hat{\cal H}_+$ which satisfy
\begin{equation}
\langle f_{\alpha}, f_{\alpha}\rangle_{M}=\langle \hat f_{\beta}, \hat f_{\beta}\rangle_{\hat M}=1\,.
\end{equation}
we can construct the amplitude 
\begin{equation}
\langle\alpha|\beta\rangle_t=\left.\langle \mu(f_{\alpha}), \hat f_{\beta}\rangle_{\hat M}\right|_t=\left.\langle f_{\alpha}, \mu^{-1}(\hat f_\beta)\rangle_M\right|_t\,,
\end{equation}
which, in general, depends on time. This gives the quantity $|\langle\alpha|\beta\rangle_t|^2$ which can be interpreted as the probability of measuring at the time $t$  the parameters  $\beta$  in the state $\alpha$ prepared on $(M,g)$ or, reversely, as the probability of measuring the parameters $\alpha$ in the state $\beta$ prepared on $(\hat M,\hat g)$. For this reason we say that $\mu(f)\in \hat{\cal K}$ is the projection of $f\in {\cal K}$. 

The isometry $\mu$ is complicated since this involves the coordinate transformation 
$\hat x=\chi(x)$ but which can be eliminated by choosing the same coordinates for the  both manifolds under consideration by taking $\chi=id \to \hat x=x$. Note that this is possible since we assumed that the local charts of $(M,g)$ and $(\hat M,\hat g)$ are included in the same domain of the flat model.  With this choice the isometry takes the simple form
\begin{equation}
\mu(f)=\left(\frac{ g^{00}\sqrt{g}}{\hat g^{00}\sqrt{\hat g}}\right)^{\frac{1}{2}}f\,,
\end{equation}
that can be used in applications.

An important particular case is when $(\hat M,\hat g)$ is just the Minkowski spacetime which is the flat model of $(M,g)$. Then we can set at any time $\chi=id$ and, in addition, we get the opportunity of defining  in $(M,g)$ states in which one measures exclusively Minkowskian parameters at a given time $t_0$. Thus for  any normalized mode function  $\hat f \in \hat{\cal K}$ on the Minkoeski spacetime we may define  the corresponding {\em Mikowskian state} on $(M,g)$ whose normalized mode function $f\in {\cal K}$ is defined such that the functions,
\begin{equation}\label{def}
\mu(f)=\left(g^{00}\sqrt{g}\right)^{\frac{1}{2}}f\,,
\end{equation}
and $\hat f$   have a {\em contact} of order $k$ at the time $t_0$, satisfying the system of $k+1$ algebraic equations,
\begin{eqnarray}
\mu(f)(t_0)&=&\hat f(t_0)\,,\nonumber\\
\frac{d\mu(f)}{dt}(t_0)&=&\frac{d\hat f}{dt}(t_0)\,,\label{Q}\\
&\vdots& \nonumber\\
\frac{d^k\mu(f)}{dt^k}(t_0)&=&\frac{d^k\hat f}{dt^k}(t_0)\,,\nonumber
\end{eqnarray}
able to give all the integration constants of $f$ in terms of the Minkowskian parameters of the function $\hat f$ we chose.  Obviously, the number $k+1$ of equations we may use depends on the number of the undetermined integration constants or other parameters we need to find out. With this method we can apply the definitions of Minkowskian particles or antiparticles to any manifold $(M,g)$ but only at a given time since, in general,  these states are evolving in time.       

\section{Rest frame vacua}

Let us consider now the family of $(1+3)$-dimensional  spatially flat FLRW spacetimes for which we use the {\em same} coordinates of the FLRW chart, $\{t,\vec{x}\}$, i. e. the proper (or cosmic) time $t\in D_t$ and the Cartesian space coordinates $\vec{x}=(x^1,x^2,x^3)\in {\Bbb R}^3$.  We denote by $M$ the spacetime whose  line element depends  on the scale factor $a(t)$ which is assumed to be a smooth function on $D_t$ giving  the of conformal time
\begin{equation}
t_c=\int \frac{dt}{a(t)}\in D_{t_c}\,,\end{equation} 
of the conformal chart $\{t_c,\vec{x}\}$. The line elements of these charts are 
\begin{equation}
ds^2=dt^2-a(t)^2 d\vec{x}\cdot d\vec{x}=a(t_c)^2\left(dt^2- d\vec{x}\cdot d\vec{x}\right)\,,
\end{equation}
where we denoted $a(t_c)=a[t(t_c)]$.  The Minkowski spacetime, denoted from now simply as $\hat M$, is the particular case when $a(t)=1$ and $t_c=t$. 

In the chart $\{t,\vec{x}\}$ the massive scalar field $\Phi : M\to {\Bbb C}$ of mass $m$ satisfies the Klein-Gordon equation 
\begin{equation}
\left(\partial_t^2+\frac{3\dot{a}(t)}{a(t)}\,\partial_t-\frac{1}{a(t)^2}\,\Delta+m^2\right)\Phi(t,\vec{x})=0\,,
\end{equation}
which allows a system of plane wave solutions, i. e. eigenfunctions of the momentum operators ${P}_i=-i\partial _i$ corresponding to the eigenvalues $(p_1,p_2,p_3)$ 
 representing  the components of the conserved momentum $\vec{p}$. These mode functions can be written as 
\begin{equation}\label{fp}
f_{\vec{p}}(t,\vec{x})=\frac{e^{i \vec{x}\cdot \vec{p}}}{[2\pi a(t)]^{\frac{3}{2}}}{\cal F}_p(t)\,,
\end{equation}
in terms of the time modulation functions  ${\cal F}_p: D_t\to {\Bbb C}$  which depend on $p=|\vec{p}|$ satisfying the equation
\begin{equation}\label{KGred}
\left[\frac{d^2}{dt^2}+\frac{p^2}{a(t)^2}+m^2-\frac{3}{2}\frac{\ddot{a}(t)}{a(t)}-\frac{3}{4}\frac{\dot{a}(t)^2}{a(t)^2}\right] {\cal F}_p(t)=0\,.
\end{equation}
which does not determine completely the form of the functions ${\cal F}_p$, remaining with integration constants which have to be fixed by supplemental assumptions. 

The fundamental solutions (\ref{fp}) form an orthonormal basis with respect to the scalar product (\ref{SPgen}) that now reads
\begin{eqnarray}
\langle f,f'\rangle_M&=&i\int_{{\Bbb R}^3} d^3x\, a(t)^3\, ( f^*\stackrel{\leftrightarrow}{\partial}_{t}f')\nonumber\\
&=&i\int_{{\Bbb R}^3} d^3x\, a(t_c)^2\, ( f^*\stackrel{\leftrightarrow}{\partial}_{t_c}f')  \,,\label{SP}
\end{eqnarray}
allowing us to impose the  normalization condition
\begin{equation}
\delta^3(\vec{p}-\vec{p}')=\langle f_{\vec{p}},f_{\vec{p}'}\rangle_M=\delta^3(\vec{p}-\vec{p}') i\,{\cal F}_p^*(t)\stackrel{\leftrightarrow}{\partial}_{t}{\cal F}_p(t)\,,
\end{equation}
requiring the time modulation functions to satisfy 
\begin{equation}\label{normF}
 i\,{\cal F}_p^*(t)\stackrel{\leftrightarrow}{\partial}_{t}{\cal F}_p(t)=1\,.
\end{equation}
Then the Klein-Gordon field can be expanded as
\begin{equation}
\Phi(x)=\int d^3p\, \left[f_{\vec{p}}(x){\frak a}(\vec{p})+f^*_{\vec{p}}(x){\frak a}^c(\vec{p})^{\dagger}\right]\,,
\end{equation}
in terms of particle, ${\frak a}, {\frak a}^{\dagger}$, and antiparticle, ${\frak a}^c, {\frak a}^{c\,\dagger}$, field operators which satisfy the canonical commutation relations
\begin{eqnarray}
\left[{\frak a}(\vec{p}),{\frak a}(\vec{p}')^{\dagger}\right]&=&\delta^3(\vec{p}-\vec{p}')\,,\\  \left[{\frak a}^c(\vec{p}),{\frak a}^c(\vec{p}')^{\dagger}\right]&=&\delta^3(\vec{p}-\vec{p}')\,.
\end{eqnarray}  

In the particular case of the Minkowski spacetime $\hat M$ the mode functions of positive frequencies of a scalar field of mass $\hat{m}$ 
\begin{equation}\label{fp0}
\hat f_{\vec{p}}(t,\vec{x})=\frac{e^{i \vec{x}\cdot \vec{p}}}{[2\pi ]^{\frac{3}{2}}}\,\hat{\cal F}_p(t)\,,\quad \hat{\cal F}_p(t)=\frac{1}{\sqrt{2 E}}\,e^{-i E t}\,,
\end{equation}
are eigenfunctions of the energy operator $i\partial_t$ depending on the conserved energy $E=\sqrt{p^2+\hat{m}^2}$ and satisfying the orthonormalization condition  with respect to the scalar product
\begin{equation}\label{SP0}
\langle\hat  f,\hat f'\rangle_{\hat M}=i\int_{{\Bbb R}^3} d^3x\, \hat f^*\stackrel{\leftrightarrow}{\partial}_{t}\hat f' \,.
\end{equation} 
On the other hand, we have shown that in any FLRW spacetime  there exists an {\em energy operator}  that in the FLRW chart, $\{t,\vec{x}\}$, has the form \cite{CSchr,CGRG} 
\begin{equation}
H=i\partial_t+\frac{\dot a(t)}{a(t)}\,\vec{x}\cdot\vec{P}\,.
\end{equation}
In general, this operator does not commute with the momentum operator $\vec{P}$ but  in the rest frames (where $\vec{p}=0$) this coincides with the Minkowski suggesting us to determine the integration constants of the solutions (\ref{fp}) by separating the frequencies just in such frames by using the Minkowskian rest states on $M$ defined in the previous section. Thus we may set the r.f.v. of the Klein-Gordon field on the FLRW manifold under consideration.

Without introducing new notations we suppose that now the mode functions (\ref{fp}) are the Minkowskian states in which  one measures in the rest frame, at the time $t_0$, the parameters of the mode functions (\ref{fp0}) for $p\to 0$ but with another rest energy, $\hat{m}\not=m$, we call here the {\em dynamical mass}. Therefore, we may consider the system (\ref{Q}) with $k=2$ giving the following equations  
\begin{eqnarray}
\lim_{p\to 0}\left.\left[{\cal F}_p(t)-\hat{\cal F}_p(t)\right]\right|_{t=t_0}&=&0\,,\label{V1}\\
\lim_{p\to 0}\left.\frac{d}{dt}\left[{\cal F}_p(t)-\hat{\cal F}_p(t)\right]\right|_{t=t_0}&=&0\,,\label{V2}\\
\lim_{p\to 0}\left.\frac{d^2}{dt^2}\left[{\cal F}_p(t)-\hat{\cal F}_p(t)\right]\right|_{t=t_0}&=&0\,.\label{V3}
\end{eqnarray}
which are enough for separating the frequencies in the rest frame and finding the dynamical mass $\hat{m}(t_0)$. Thus the first two equations give the normalized integration constants corresponding to the r.f.v. while the third one  helps us to find the associated dynamical mass in the rest frame. All these quantities may depend on the time $t_0$ when we impose the Minkowskian form of the mode functions in the rest frame. This means that, in general,  the r.f.v. is dynamic, being  associated with a time-dependent dynamical mass. Nevertheless,  this vacuum becomes stable on the FLRW manifolds where the energy operator is conserved, i. e. the Minkowski and de Sitter spacetimes, since then the energy operator in the rest frame, $i\partial_t$,  {\em commutes} with that of the field equation completing thus the s.c.c.o. but only in the rest frame.  
while for the functions $K_{\nu}$ we have to use Eq. (\ref{IK}).

\section{Applications}

For solving concrete examples we may start with a  time modulation function of the general form 
\begin{equation}\label{Fphi}
{\cal F}_p(t)=c_1\phi_p(t)+c_2\phi^*_p(t)\,,
\end{equation}
where $\phi_p$ is a particular solution satisfying 
\begin{equation}
\left(\phi_p, \phi_p\right)=1 ~~\to~~\left(\phi^*_p, \phi^*_p\right)=-1\,.
\end{equation}
The normalized solutions of positive frequency, $f_{\vec{p}}\in {\cal K}_+$,  must have  time modulation functions which satisfy
\begin{equation}
 \left({\cal F}_p , {\cal F}_p\right)=1~~\to~~|c_1|^2-|c_2|^2=1\,.
\end{equation} 

In the rest frame (where $p=0$) we denote simply $\phi=\phi_p|_{p=0}$  such that the system  (\ref{V2}) can be written as
\begin{eqnarray}
c_1\phi(t_0)+c_2\phi^*(t_0)&=&\frac{1}{\sqrt{2 \hat m}}e^{-i\hat m t_0}\,,\\
c_1\dot{\phi}(t_0)+c_2\dot{\phi}^*(t_0)&=&-i\hat m\frac{1}{\sqrt{2 \hat m}}e^{-i\hat m t_0}\,,\\
c_1\ddot{\phi}(t_0)+c_2\ddot{\phi}^*(t_0)&=&-\hat m^2\frac{1}{\sqrt{2 \hat m}}e^{-i\hat m t_0}\,.
\end{eqnarray}
The first two equations give the normalized integration constants corresponding to the r.f.v.,
\begin{eqnarray}
c_1\to c_1(t_0)&=&\frac{e^{-i\Omega(t_0)t_0}}{\sqrt{2\Omega(t_0)}}\left(\Omega(t_0)\phi^*(t_0)-i\dot \phi^*(t_0)\right)\,,~~~\label{c1t}\\
c_2\to c_2(t_0)&=&\frac{e^{-i\Omega(t_0)t_0}}{\sqrt{2\Omega(t_0)}}\left(-\Omega(t_0)\phi(t_0)+i\dot \phi(t_0)\right)\,,~~~\label{c2t}
\end{eqnarray}
while the third one gives us the associated dynamical mass in the rest frame, 
\begin{equation}
\hat m\to \hat m(t_0)=\lim_{p\to 0}\Omega_p(t_0)\equiv \Omega(t_0)\,,
\end{equation} 
since $\ddot \phi=-\Omega^2 \phi$ as in Eq. (\ref{KGred}). 

Thus we find that a particle prepared  in r.f.v. at the time $t_0$ has the mode function
\begin{equation}\label{fp1}
f_{\vec{p},t_0}(t,\vec{x})=\frac{e^{i \vec{x}\cdot \vec{p}}}{[2\pi a(t)]^{\frac{3}{2}}}{\cal F}_p(t_0,t)\,,
\end{equation}
whose time modulation function
\begin{equation}\label{Fptt} 
{\cal F}_p(t_0,t)=c_1(t_0)\phi_p(t)+c_2(t_0)\phi^*_p(t)\,.
\end{equation}
depends on the integration constants (\ref{c1t}) and (\ref{c2t}) which comply with the normalization condition
\begin{equation}
|c_1(t_0)|^2-|c_2(t_0)|^2=\left\{
\begin{array}{lll}
1&{\rm if}& \Omega(t_0)^2>0\\
0&{\rm if}& \Omega(t_0)^2<0
\end{array}\right.\,.
\end{equation} 
The set $\{f_{\vec{p},t_0}| \vec{p}\in {\Bbb R}^3\}$ forms a basis in ${\cal K}_+$ while the set $\{f_{\vec{p},t_0}^*| \vec{p}\in {\Bbb R}^3\}$ is the corresponding basis of  ${\cal K}_-$ in the r.f.v. prepared at $t=t_0$.

In general,  the r.f.v. is dynamic, being  associated with a time-dependent dynamical mass $\hat m(t)=\Omega(t)\in {\Bbb R}$. The time domain $D_t=D_t^+\cup D_t^-$ is split into the tardyonic part $D_t^+=\{t|\Omega(t)^2>0\}$ and the tachyonic one,  $D_t^-=\{t|\Omega(t)^2<0\}$.  All the tachyonic states with $\Omega(t)=i|\Omega(t)|$ are eliminated as having null norms. Thus in r.f.v. the scalar field survives only on $D_t^+$.

As mentioned the r.f.v. becomes stable only on the Minkowski and de Sitter spacetimes where the energy operator is conserved,  satisfying $[H_0,\Omega]=i\partial_t\Omega=0$.
  
\subsection{de Sitter expanding universe}

Let us consider first an example of stable r.f.v. on the expanding portion of the de Sitter spacetime, $M$, having the scale factor  $a(t)=e^{2\omega t}$ (where  $\omega$ is the Hubble de Sitter constant in our notation) defined for $t\in(-\infty,\infty)$,  giving the conformal time $t_c$ and the function $a(t_c)$ as
\begin{equation}\label{tc}
t_c=-\frac{1}{\omega} e^{-\omega t}\in (-\infty,0]\,, \quad a(t_c)=-\frac{1}{\omega t_c}\,.
\end{equation}
In the conformal chart the Klein-Gordon equation is analytically solvable giving the mode functions of the momentum basis of the form (\ref{fp}) having the time modulation functions
\begin{equation}\label{fdS}
{\cal F}_p(t_c)=\ c_1\phi_p(t) + c_2\phi_p^*(t)\,,\quad \phi_p(t)=\frac{1}{\sqrt{\pi\omega}}K_{\nu}(ipt_c)\,,
\end{equation}
where
\begin{equation}\label{ndS}
\nu=\left\{\begin{array}{lll}
\sqrt{\frac{9}{4}-\mu^2}&{\rm for} & \mu<\frac{3}{2}\\
i\kappa\,,\quad \kappa= \sqrt{\mu^2-\frac{9}{4}}&{\rm  for} & \mu>\frac{3}{2}
\end{array} \right. \,, \quad \mu=\frac{m}{\omega}\,.
\end{equation}
By using  Eq. (\ref{KuKu}) we find that the normalization condition (\ref{normF}) is fulfilled only if we take
\begin{equation}\label{norC}
\left|c_1\right|^2-\left|c_2\right|^2=1\,.
\end{equation}

We  assume first that $m>\frac{3}{2}\,\omega$ solving the system (\ref{V2}) in the conformal chart $\{t_c,\vec{x}\}$ where the de Sitter time modulation function has the form (\ref{fdS}) with  $\nu=i\kappa$ while the Minkowski one (\ref{fp0}) takes the form
\begin{equation}
\hat{\cal F}[t(t_c)]=\frac{(-\omega t_c)^{\frac{i E}{\omega}}}{\sqrt{2E}}\,.
\end{equation}
Moreover, since in this case the limit to $p\to 0$ is sensitive, we solve first this system for $p\not= 0$ and then we evaluate this limit. From the first two equations we obtain the integration constants
\begin{eqnarray}
c_1(p)&=&\frac{(-\omega t_c)^{\frac{i E}{\omega}}}{\sqrt{2 \pi\omega E}}\left[\omega p t_c K_{i\kappa+1}(-ipt_c)\right.\nonumber\\
&&\hspace*{14mm}\left.+(E-\kappa\omega)K_{i\kappa}(-ipt_c)\right]\,,\\
c_2(p)&=&-\frac{(-\omega t_c)^{\frac{i E}{\omega}}}{\sqrt{2 \pi\omega E}}\left[\omega p t_c K_{i\kappa+1}(ipt_c)\right.\nonumber\\
&&\hspace*{14mm}\left.+(E-\kappa\omega)K_{i\kappa}(ipt_c)\right]\,,
\end{eqnarray}  
while from the last one, 
\begin{equation}
\lim_{p\to 0}\left[E^2-\kappa^2\omega^2-\omega^2p^2 t_c^2\right]=(\hat m^2-\omega^2\kappa^2)=0\,,
\end{equation}
gives the expected dynamical mass 
\begin{equation}\label{dm}
\hat m=\omega\kappa=\sqrt{m^2-\frac{9}{4}\,\omega^2}\,,
\end{equation}
related to the well-known rest energy \cite{CGRG}.  Then for  $p\to 0$  we obtain the constants  $c_1=\lim_{p\to0}c_1(p)$ and $c_2=\lim_{p\to0}c_2(p)$ which have the absolute values   
\begin{eqnarray}
|c_1|&=&\frac{e^{\pi\kappa}}{\sqrt{e^{2\pi\kappa}-1}}\,,\label{C1}\\
|c_2|&=& \frac{1}{\sqrt{e^{2\pi\kappa}-1}}\,,\label{C2}
\end{eqnarray}  
resulted from Eqs. (\ref{IK}) and   (\ref{I0}).   Finally, by substituting these values in Eq. (\ref{fdS}), we obtain the definitive result of the time modulation functions of positive energy in the r.f.v.,
\begin{equation}
{\cal F}_p(t_c)=\sqrt{\frac{\pi}{\omega}}\left(\frac{p}{2\omega}\right)^{-i\kappa}\frac{I_{i\kappa}(ipt_c)}{\sqrt{e^{2\pi\kappa}-1}}\,,
\end{equation}
where the general phase factor was introduced for assuring the correct limit for $p\to 0$ as given by Eq. (\ref{I0}). These functions are correctly normalized since the integration constants (\ref{C1}) and (\ref{C2}) satisfy the condition (\ref{norC}). Note that these results can be rewritten in terms of the cosmic time $t$ according to Eq. (\ref{tc}).

Furthermore, we consider the case of $m<\frac{3}{2}\,\omega$ applying the same method for fixing the r.f.v.. We solve first two equations of the system  (\ref{V2}) for $p\not=0$ and an arbitrary time $t_c$ obtaining
\begin{eqnarray}
c_1(p,t_c)&=&\frac{(-\omega t_c)^{\frac{i E}{\omega}}}{\sqrt{2 \pi\omega E}}\left[(E+i\nu\omega)K_{\nu}(-ipt_c)\right.\nonumber\\
&&\hspace*{18mm}\left. -\omega p t_c K_{\nu+1}(-ipt_c)\right]\,, \\
c_2(p,t_c)&=&\frac{(-\omega t_c)^{\frac{i E}{\omega}}}{\sqrt{2  \pi\omega E}}\left[(E+i\nu\omega)K_{\nu}(ipt_c)\right.\nonumber\\
&&\hspace*{18mm}\left. +\omega p t_c K_{\nu+1}(ipt_c)\right]\,.
\end{eqnarray} 
From the thisrd equation we find the expected condition
\begin{equation}
\lim_{p\to 0}\left[E^2+\nu^2\omega^2-\omega^2p^2 t_c^2\right]=(\hat m^2+\omega^2\nu^2)=0\,,
\end{equation}
giving the tachyonic dynamical mass $\hat m=\pm i\nu\omega$. Moreover, we find that in the rest frame we have
\begin{equation}
\lim_{p\to 0} c_1(p,t_c)=\lim_{p\to 0} c_2(p,t_c)=0\,,
\end{equation} 
which means that if we set the r.f.v. then the particles with $m<\frac{3}{2}\,\omega$  cannot survive on the de Sitter expanding portion. 

The above results can be now gathered in the synthetic form of the mode functions of positive frequency in the conformal chart,  
\begin{equation}
f_{\vec{p}}(t_c,\vec{x})=\left(\frac{-\omega t_c}{2\pi}\right)^{\frac{3}{2}}\sqrt{\frac{\pi}{\omega}}\left(\frac{p}{2\omega}\right)^{-\nu}\frac{I_{\nu}(ipt_c)\,e^{i\vec{p}\cdot\vec{x}}}{\sqrt{e^{-2i\pi\nu}-1}}\,,
\end{equation} 
that hold for any real or imaginary value of  $\nu$, given by Eq. (\ref{ndS}). In the tachyonic case, when $\nu$ takes real values, the squared norm of $f_{\vec{p}}$ vanishes since then we have $ I_{\nu}(-i pt_c) \stackrel{\leftrightarrow}{\partial_{t_c}}I_{\nu}(ipt_c)\propto I_{\nu}(i pt_c) \stackrel{\leftrightarrow}{\partial_{t_c}}I_{\nu}(ipt_c)= 0$. 

Thus we have shown that the scalar r.f.v. on the de Sitter expanding universe is stable  corresponding to a time-independent dynamical mass (\ref{dm}) which does make sense  only  when  $m>\frac{3}{2}\,\omega$. In other words, the frequencies separation in the rest frames can be done only for the scalar fields which satisfy this condition. 

Otherwise we have either to eliminate the scalar fields with $m<\frac{3}{2}\,\omega$  or to resort to another vacuum as the adiabatic Bunch-Davies  one \cite{BD} which can be set for particles of any mass by taking $c_1=\frac{1}{\sqrt{\pi\omega}}$ and $c_2=0$.

\subsection{Milne-type spatially flat FLRW spacetime}

Let us consider now an example of manifold $M$ where we do not have adiabatic vacua remaining only with an unstable r.f.v. corresponding to a time-dependent dynamical mass.  This is the $(1+3)$-dimensional spatially flat FLRW manifold  with the scale factor $a(t)=\omega t$ determining  the conformal time as
\begin{equation}\label{tt}
t_c=\int \frac{dt}{a(t)}=\frac{1}{\omega} \ln(\omega t)\in  (-\infty,\infty)~\to~  a(t_c)=e^{\omega t_c}\,.
\end{equation}
The constant $\omega$, introduced from dimensional considerations, is an useful free parameter which in the case of the genuine Milne's universe (of negative space curvature)  must be fixed to $\omega=1$ for eliminating the gravitational sources \cite{BD}.

This spacetime $M$ is produced by isotropic gravitational sources, i. e. the density $\rho$ and pressure $p$, evolving in time as
\begin{equation}
\rho=\frac{3}{8\pi G}\frac{1}{t^2}\,, \quad p=-\frac{1}{8\pi G}\frac{1}{t^2}\,,
\end{equation}
and vanishing for $t\to\infty$. These sources govern the expansion of $M$ that can be better observed in the chart $\{t, \vec{\hat x}\}$, of 'physical' space coordinates $\hat x^i=\omega t x^i$, where the line element 
 \begin{equation}
 ds^2=\left(1-\frac{1}{t^2}\vec{\hat x}\cdot \vec{\hat x}\right)dt^2 + 2 \vec{\hat x}\cdot d\vec{\hat x}\,\frac{dt}{t}-d\vec{\hat x}\cdot d\vec{\hat x}\,,
 \end{equation}
lays out an expanding horizon at $|\vec{\hat x}|=t$ and tends to the Minkowski spacetime when $t\to \infty$ and the gravitational sources vanish.

In the FLRW chart $\{t,\vec{x}\}$ of this spacetime the Klein-Gordon equation is analytically solvable, the fundamental solutions  having the time modulation functions 
\begin{equation}\label{solM}
{\cal F}_p(t)=c_1\phi_p(t) +c_2\phi_p^*(t)\,,\quad \phi_p(t)=\sqrt{\frac{t}{\pi}}\, K_{\nu}(imt)\,,
\end{equation} 
where 
\begin{equation}\label{nM}
\nu=\sqrt{1-\frac{p^2}{\omega^2}}\,,
\end{equation}
can take real or pure imaginary values for $p>\omega$. 

{ \begin{figure}
\centering
  \includegraphics[scale=0.70]{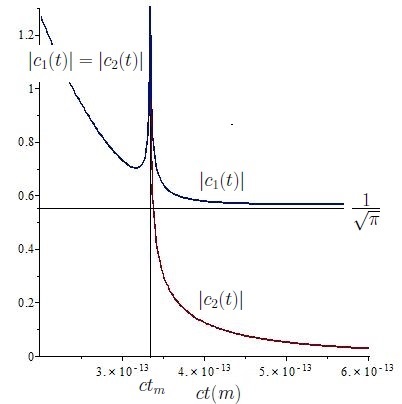}
  \caption{The functions $|c_1(t)|$ and $|c_2(t)|$  versus $ct$ for a light particle having the electron mass, $m=m_e$, for which $t_{m_e}\sim 1.1\, 10^{-21} s$. The plotting domain is $0.6\, t_{m_e}<t< 1.8\, t_{m_e}$. }
 \end{figure}}

We observe that here we cannot speak about adiabatic vacua as long as the functions (\ref{solM}) are singular in $t=0$. Therefore, we must focus only  the r.f.v. for which the time-dependent integration constants,
\begin{eqnarray}
c_1(t)&=&\frac{e^{-it\hat m}}{2 \sqrt{2 t\hat m}}\left[2t mK_0(-imt)\right. \nonumber\\
&&~~~~~~~~~~+\left.(i+2t\hat m))K_1(-imt) \right]\,,\label{CM1}\\
c_2(t)&=&\frac{e^{-it\hat m}}{2 \sqrt{2 t\hat m}}\left[2t mK_0(imt)\right.\nonumber\\
&&~~~~~~~~~~-\left.(i+2t\hat m))K_1(imt) \right]\,,\label{CM2}
\end{eqnarray}
result from Eqs. (\ref{c1t}) and (\ref{c2t}).
The corresponding dynamical mass reads
\begin{equation}\label{mM}
\hat m(t)=\sqrt{m^2-\frac{3}{4\,t^2}}\,.
\end{equation}
The functions (\ref{CM1}) and (\ref{CM2}) are singular in $t=0$ and  $t=t_m\equiv\frac{\sqrt{3}}{2 m}$ when $\hat m(t)$ vanishes (as in Fig. 1). From Eq. (\ref{mM})  we see  that a particle of mass $m$ has a tachyonic behavior in the domain $D_t^-=(0,t_m)$ and a tardyonic one  only if $t\in D_t^+=(t_m, \infty)$. As in the general case, we can verify that 
\begin{equation}
|c_1(t)|^2-|c_2(t)|^2=\left\{
\begin{array}{lll}
0&{\rm if}&0< t<t_m\\
{1}&{\rm if}&t>t_m
\end{array}\right.
\end{equation} 
showing that on the tachyonic domain the wave function is of null norm having thus no physical meaning. 

This means that the scalar particles can be prepared only  in the tardyonic domain $t>t_m$ where $\hat m(t)$ increases with $t$ such that for $t\to\infty$, when $M$ becomes just the Minkowski spacetime, this tends to $m$. Moreover, in this limit we recover the usual Minkowski scalar modes since the functions $K$ behave as in Eq. (\ref{Km0}) such that
\begin{equation}\label{limc1c2}
\lim_{t\to\infty}|c_1(t)|=1\,, \quad \lim_{t\to\infty}|c_2(t)|=0\,.
\end{equation}

All these results can be encapsulated in the definitive form of the mode functions of positive frequency, prepared at the time $t_0>t_m$ and defined for $t>t_0$, that read
\begin{eqnarray}
f_{\vec{p}}(t_0,t,\vec{x})&=&\frac{e^{i \vec{p}\cdot\vec{x}}}{(2\pi\omega t)^{\frac{3}{2}}}\left[c_1(t_0) \sqrt{\frac{t}{\pi}}\, K_{\nu}(imt)\right.\nonumber\\
&&\hspace*{12mm}\left.+c_2(t_0) \sqrt{\frac{t}{\pi}}\, K_{\nu}(-imt)\right]\,,
\end{eqnarray}
where $\nu$ depends on $p$ as in Eq.  (\ref{nM}).

{ \begin{figure}
\centering
  \includegraphics[scale=0.65]{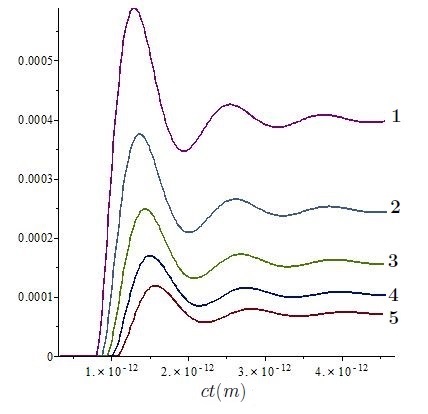}
  \caption{The function $n(t_0,t)$ versus $ct$ in the domain $t_m<t<14\, t_m$ for $m=m_e$ and: $t_0=2.4\, t_m$ (1), $t_0=2.6\, t_m$ (2), $t_0=2.8\, t_m$ (3), $t_0=3\, t_m$ (4), $t_0=3.2\, t_m$ (5).}
 \end{figure}}

The instability of r.f.v. on this expanding manifold give rise to c.p.c. that can be analyzed thanks to our  previous results that hold for any $t>t_0$. Thus we can study how the particles created at $t_0$ can be measured at any moment $t>t_0$ calculating the Bogolyubov coefficients between the bases $f_{\vec{p}}(t_0)$ and  $f_{\vec{p}}(t)$ which, according to Eq. (\ref{KuKu}), read
\begin{eqnarray}
\alpha(\vec{p},t_0;\vec{p}',t)&=&\langle f_{\vec{p}}(t_0), f_{\vec{p}'}(t)\rangle=\delta^3(\vec{p}-\vec{p}') \nonumber\\
&\times&\left[c_1^*(t_0)c_1(t)-c_2^*(t_0)c_2(t) \right]\,, \\
\beta(\vec{p},t_0;\vec{p}',t)&=&\langle f_{\vec{p}}^*(t_0), f_{\vec{p}'}(t)\rangle=\delta^3(\vec{p}-\vec{p}') \nonumber\\
&\times& \left[c_2(t_0)c_1(t)-c_1(t_0)c_2(t) \right]\,, 
\end{eqnarray}  
Then the density of the new particles or antiparticles created between $t_0$ and $t$ is proportional to,
\begin{equation}\label{dens}
n(t_0,t)\propto\left| c_2(t_0)c_1(t)-c_1(t_0)c_2(t)\right|^2_,.
\end{equation}

\noindent In addition, we observe that the rate of c.p.c. can also be estimated as
\begin{equation}\label{Rate} 
R(t_0,t)\propto\frac{d\, n(t_0,t)}{dt}\,. 
\end{equation} 
Thus we can point out the effects of the dynamic r.f.v. which tends to stability when the time is increasing since then
\begin{equation}
\lim_{t\to \infty} n(t_0,t)\sim |c_2(t_0)|^2\,, \quad \lim_{t\to \infty} R(t_0,t)=0\,,
\end{equation}
as we deduce from Eqs. (\ref{limc1c2}). Hereby we see that the dynamical effect is visible only for the very old particles, prepared at $t_0<5\,t_m$, since the function $c_2(t_0)$ decreases rapidly to zero when $t_0$ increases and $\hat m(t_0)\to m$. Thus for the younger particles, prepared at $t_0 > 5-10\, t_m$, the dynamical effect is inhibited remaining with an apparently stable r.f.v. of the Bunch-Davies type (with $c_1=1$ and $c_2=0$)  in which the mode functions can be approximated as
\begin{equation}
f_{\vec{p}}(t,\vec{x})\sim \frac{e^{i \vec{p}\cdot\vec{x}}}{(2\pi\omega t)^{\frac{3}{2}}}\sqrt{\frac{t}{\pi}}\,  K_{\nu}(imt)\,,
\end{equation}  
independent on the moment $t_0$ when the particle was prepared.

{ \begin{figure}
\centering
  \includegraphics[scale=0.65]{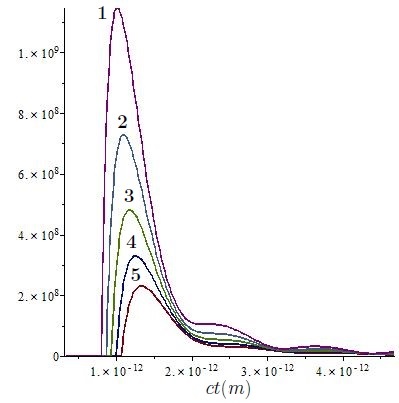}
  \caption{The function $R(t_0,t)$ versus $ct$ in the domain $t_m<t<14\, t_m$ for $m=m_e$ and: $t_0=2.4\, t_m$ (1), $t_0=2.6\, t_m$ (2), $t_0=2.8\, t_m$ (3), $t_0=3\, t_m$ (4), $t_0=3.2\, t_m$ (5).}
 \end{figure}}

Finally we must specify that, in general, the dynamic effect discussed above is very fast, during an extremely  short period of time, even at quantum scale, since by definition $t_m=\frac{\sqrt{3}}{2m}$ (or $\frac{\sqrt{3}}{2}\frac{\hbar}{m c^2}$ in SI units)  is very small. For example, if we take $m$ to be just the electron mass $m_e$ then $t_{m_e} \sim 1.1\, 10^{-21} s$ such that for the particles born at cosmic times  $t_0>10^{-20} s$  the r.f.v.  is apparently stable.  Only the particles prepared at $t_0<10^{-20} s$ lay out this effect as it results from  Figs. 2 and 3 where we plot the functions (\ref{dens}) and (\ref{Rate}) versus $ct$ instead of $t$ for avoiding too small numbers. Thus it is obvious that the dynamical effects of the r.f.v. may be of interest only at quantum scale in the cosmology of the very early Milne-type universe.

\section{Concluding remarks}

We proposed here a method of projecting the quantum states from a state space of a given geometry into another state space generated by a different geometry, keeping the correct normalization which is crucial in interpreting the quantum quantities (probabilities, expectation values, transition amplitudes, etc.). This method helped us to define, on any spatially flat FLRW  spacetime, the Minkowskian states we need for setting the r.f.v. of the massive scalar field which, in contrast to the Dirac one, does not have a Minkowskian behavior in rest frames on the FLRW manifolds.  In this manner, we obtained a stable r.f.v. on the de Sitter expanding universe and, for the first time, we found a dynamical vacuum, corresponding to a time-dependent dynamical mass on a Milne-type spacetime. In this last case, the dynamic r.f.v.  gives rise to a very fast c.p.c. that could be of interest but only in the very early Milne-type universe.  It is remarkable that in r.f.v. all the possible tachyonic behaviors (e.g. for $m<\frac{3}{2}\,\omega$ in the de Sitter case and $t<t_m$ in the  Milne-type universe) are eliminated in a natural manner, the corresponding mode functions resulting to have null norms. These results may improve the study of c.p.c. on the FLRW manifolds combining the r.f.v. with the other vacua proposed so far.

On the other hand, we must stress that the r.f.v. cannot be defined for the massless fields which do not have rest frames. In the case of the Maxwell and massless  Dirac fields this is not an impediment since the neutrino and Maxwell equations are confomally covariant such that in the conformal charts of the FLRW spacetimes  one may take over the frequency separation from the flat case. The only problem which remains partially unsolved  is the vacuum  of the massless scalar field whose equation is no longer covariant under conformal transformations.  This sensitive case is revisited time by time with the hope of finding a convenient interpretation \cite{coco}. 

Another approach is the quantum theory of interacting fields on curved manifolds in which the amplitudes of the quantum transitions can be calculated by using perturbations in terms of free fields \cite{Lot1,Lot2,Lot3,R1,R2,R3,A1,A2} as in our recent de Sitter QED \cite{CQED,Cr1,Cr2}. Even though in this framework only adiabatic vacua were considered so far, we have now the opportunity of using many types of vacua for improving the calculation of the transition amplitudes. Thus, for example,  in a collision process we may take the incident beam in the adiabatic vacuum and the target in the r.f.v.. Moreover,  for the internal lines of the Feynman diagrams the r.f.v. is the favorite candidate since this can be defined naturally for the massive fields on any spatially flat FLRW spacetime. Thus by using many well-defined vacua we could combine the methods of c.p.c. with those of the perturbative quantum field theory for analysing various quantum effects in evolving universes.

\appendix

\setcounter{section}{0}\renewcommand{\thesection}{\Alph{section}}
\setcounter{equation}{0} \renewcommand{\theequation}
{A.\arabic{equation}}
\section{Modified Bessel functions}

The modified Bessel functions $I_{\nu}(z)$ and $K_{\nu}(z)$ are related as \cite{NIST}
\begin{eqnarray}
K_{\nu}(z)&=&K_{-\nu}(z)=\frac{\pi}{2}\frac{I_{-\nu}(z)-I_{\nu}(z)}{\sin\pi \nu}\,,\label{IK}\\
I_{\pm\nu}(z)&=&e^{\mp i\pi\nu}I_{\pm\nu}(-z)\nonumber\\
&=&\frac{i}{\pi}\left[K_{\nu}(-z)-e^{\mp i\pi\nu}K_{\nu}(z)\right]\,.\label{KI}
\end{eqnarray}
Their Wronskians  give  the identities we need for normalizing the mode functions. For $\nu=i\mu$ we obtain
\begin{equation}
i I_{i\mu}(i s) \stackrel{\leftrightarrow}{\partial_{s}}I_{-i\mu}(is)= \frac{2\, {\rm sinh}\,\pi\mu}{\pi s}\,,\label{IuIu} 
\end{equation}
while the identity
\begin{equation}
i K_{\nu}(-i s) \stackrel{\leftrightarrow}{\partial_{s}}K_{\nu}(is)
=\frac{\pi}{|s|}\,,\label{KuKu}
\end{equation}
holds for any $\nu$.

For $|z|\to \infty$ and any $\nu$ we have,
\begin{equation}\label{Km0}
 I_{\nu}(z) \to \sqrt{\frac{\pi}{2z}}e^{z}\,, \quad K_{\nu}(z) \to K_{\frac{1}{2}}(z)=\sqrt{\frac{\pi}{2z}}e^{-z}\,.
\end{equation} 
In the limit of $|z|\to 0$ the functions $I_{\nu}$ behave as  
\begin{equation}\label{I0}
I_{\nu}(z)\sim \frac{1}{\Gamma(\nu+1)} \left(\frac{z}{2}\right)^{\nu}\,,
\end{equation}
while for the functions $K_{\nu}$ we have to use Eq. (\ref{IK}).

\end{document}